\documentstyle[12pt,a4]{article}
\title{\sf Rare-Earth Nuclei: Radii, Isotope-Shifts and Deformation
Properties in the Relativistic Mean Field Theory}
\author{G.A. Lalazissis$^1$, M.M. Sharma$^2$ and P. Ring$^1$\\
$^1$Physik Department, Technische Universit\"at M\"unchen \\
D-85747 Garching, Germany\\
$^2$Max Planck Institut f\"ur Astrophysik,\\
D-85740 Garching bei M\"unchen, Germany}

\begin{document}

\maketitle
\begin{abstract}
A systematic study of the ground-state properties of
even-even rare earth nuclei has been performed in the
framework of the Relativistic Mean-Field (RMF) theory using the
parameter set NL-SH. Nuclear radii, isotope shifts and deformation
properties of the heavier rare-earth nuclei have been
obtained, which encompass atomic numbers ranging from Z=60
to Z=70 and include a large range of isospin. It is shown
that RMF theory is able to provide a good and
comprehensive description of the empirical binding energies
of the isotopic chains.  At the same time the quadrupole
deformations $\beta_{2}$ obtained in the RMF theory are found
to be in good agreement with the available empirical values.
The theory predicts a shape transition from prolate to
oblate for nuclei at neutron number N=78 in all the chains.
A further addition of neutrons up to the magic number 82
brings about the spherical shape. For nuclei above N=82,
the RMF theory predicts the well-known onset of prolate deformation
at about N=88, which saturates at about N=102. The deformation
properties display an identical behaviour for all the nuclear
chains. A good description of the above deformation transitions
in the RMF theory in all the isotopic chains leads to a successful
reproduction of the anomalous behaviour of the empirical
isotopic shifts of the rare-earth nuclei. The RMF theory
exhibits a remarkable success in providing a unified and
microscopic description of various empirical data.

\end{abstract}

\newpage
\baselineskip=24pt
\section{\sf INTRODUCTION}
The Relativistic Mean-Field (RMF) theory \cite{SW.86,Ser.92}
has proved to be a powerful tool for an effective
microscopic description of nuclear structure at and away
from the line of stability.  With a very limited number of
parameters in a non-linear \cite{BB.77} nuclear Lagrangian,
one is able to obtain a very good quantitative
description of the ground-state properties of spherical and
deformed nuclei \cite{GRT.90,SNR.93,LS.95} at and away from the
stability line \cite{SLR.93,SLH.94,SHR.94}. Successful attempts
have also been made to obtain dynamical properties such as
collective excitations \cite{VBR.94} and identical bands in
rotating superdeformed nuclei \cite{KR.93}.

In the RMF theory the saturation and the density dependence
of the nuclear interaction is obtained by a balance between
a large attractive scalar $\sigma$-meson field and a large
repulsive vector $\omega$-meson field. The asymmetry
component is provided by the isovector $\rho$ meson. The
nuclear interaction is hence generated by the exchange of
various mesons between nucleons in the framework of the
mean field. The spin-orbit interaction arises naturally in
the RMF theory as a result of the Dirac structure of
nucleons.

In an earlier work \cite{SLR.93} we studied the isotope
shifts of Pb nuclei. It was shown that the anomalous kink
at the shell closure (at N=126) \cite{SLR.93} could be
described within the RMF theory successfully.
This long standing problem remained a puzzle within the
non-relativistic mean-field approach using various Skyrme
type of forces \cite{TBF.93}. The spin-orbit term in the RMF
theory, which is different from that in the Skyrme
approach, lies at the origin of this success. Inspired by
the success of the RMF theory to explaining various subtle
aspects originating from shell effects, the form of the
spin-orbit potential in the Skyrme theory has been modified
and a Modified Skyrme Ansatz (MSkA) has been proposed
\cite{LSK.94,SLK.95}, which includes only the one-body (direct)
component of the spin-orbit force in the Skyrme theory
vis-a'-vis the usual Skyrme theory which includes also the
exchange term in the spin-orbit potential.  Consequently,
this has led to a success in obtaining the kink in the
isotope shifts of Pb nuclei in the MSkA. The broader
implications of the MSkA are under investigation.  Another
approach to tackling this problem of the kink in the
isotope shifts was undertaken in the framework of
conventional Skyrme theory in Ref. \cite{RF.95} by
introducing an additional parameter in the two-body spin-orbit
force.

Recently a systematic study \cite{LS.95} of ground-state
properties of Kr, Sr and Zr isotopes was performed in the
framework of the RMF theory using the force NL-SH
\cite{SLR.93}. It was shown that the RMF theory provides a
good description of the binding energies, charge radii
and deformation properties of nuclei in the Z=40 region
over a large range of isospin.  The RMF theory has predicted a
complex web of many dramatic shape transitions in the
isotopic chains of Sr, Kr and Zr. A shape coexistence in
several heavy Sr isotopes has also been seen. It was
observed that the RMF theory is able to describe the anomalous
kinks in isotope shifts in Kr and Sr nuclei, the problem
which was hitherto unsolved \cite{Ott.89,BC.95}. In this case
deformation changes play an important role to understand the
experimental isotope shifts, and a combined effect of the
theory to be able to reproduce deformation and the inherent
shell effects in the RMF theory lead to the above results.

The rare-earth nuclei have been of great experimental as
well as theoretical interest.  Most of the studies are confined
to employing the approach of Nilsson-Strutinsky \cite{AFN.90} or
using the Skyrme Ansatz for deformed nuclei \cite{Vau.73}.
Self-consistent calculations with Skyrme or Gogny forces
were performed for a number of rare-earth isotopes
\cite{Vau.73,FQV.73,Que.75,DGG.75,QF.78,Gog.75b,SLV.79,BQB.82}.
Relativistic calculations with the force NL1 \cite{Rei.89}
have also been performed for a few representative
rare-earth nuclei \cite{GRT.90}.  However, in most of these
studies only some specific rare-earth nuclei were
investigated, hindered largely by large computing time
needed for such calculations and also due to uncertainties
in the available interactions. In this investigation we
use the parameter set NL-SH, which provided a very good
description of the ground-state properties of deformed Xe
nuclei \cite{SNR.93} and of nuclei in the region of
Z=40 \cite{LS.95}.  We study systematically, the
ground-state properties of six isotopic chains of
rare-earth nuclei with atomic numbers $60 \leq Z
\leq 70$ covering a large range of masses. This is aimed
at drawing some general conclusions about the properties of
these nuclei as well as the trend of their variation with
the neutron number. We focus our attention mainly on the
sizes, the isotope shifts and the quadrupole deformations
due to the special interest they present. It is worth
noting that the isotope shifts and the deformations of the
heavier rare-earth nuclei have been measured systematically
by Neugart et al. \cite{BMS.82,Neu.82,NWA.83,Neu.85} and
also by the Lenigrad-Troitz collaboration
\cite{Alk.87,LM.87,ABD.88}. In addition experimental data are
available on the quadrupole and hexadecapole moments of
various rare-earth nuclei. (see for example
\cite{CBH.76,WWE.74} and also
\cite{Hil.70,Cli.73,LS.72,Ste.70,Ste.71,Bru.73,Say.72,Gre.73}).

In this work we present a detailed and exhaustive
microscopic study of the ground-state properties of several
isotopic chains of nuclei in the rare-earth region. In
section II we give some details of the formalism of the RMF
theory used for deformed nuclei. In section III we provide
some details of the calculations.  In section IV we
present and discuss our results. A comparison of
the RMF results is made with empirical data
wherever available. The last section summarizes our main
conclusions.

\section{\sf THE RELATIVISTIC MEAN-FIELD FORMALISM}

The starting point of the RMF theory is a Lagrangian
density \cite{SW.86,Ser.92} which describes the nucleons as Dirac
spinors interacting via the exchange of several mesons. The
Lagrangian density can be written in the following form:
\begin{equation}
\begin{array}{rl}
{\cal L} &=
\bar \psi (i\rlap{/}\partial -M) \psi +
\,{1\over2}\partial_\mu\sigma\partial^\mu\sigma-U(\sigma)
-{1\over4}\Omega_{\mu\nu}\Omega^{\mu\nu}+\\
\                                        \\
\ & {1\over2}m_\omega^2\omega_\mu\omega^\mu
-{1\over4}{\vec R}_{\mu\nu}{\vec R}^{\mu\nu} +
 {1\over2}m_{\rho}^{2} \vec\rho_\mu\vec\rho^\mu
-{1\over4}F_{\mu\nu}F^{\mu\nu} \\
\                              \\
 &  g_{\sigma}\bar\psi \sigma \psi~
     -~g_{\omega}\bar\psi \rlap{/}\omega \psi~
     -~g_{\rho}  \bar\psi
      \rlap{/}\vec\rho
      \vec\tau \psi
     -~e \bar\psi \rlap{/}A \psi
\end{array}
\end{equation}
The meson fields are the isoscalar $\sigma$ meson, the
isoscalar-vector $\omega$ meson and the isovector-vector
$\rho$ meson. The latter provides the necessary isospin
asymmetry. The arrows denote the isovector quantities. The
model contains also a non-linear scalar self-interaction of
the $\sigma$ meson :

\begin{equation}
U(\sigma)~={1\over2}m_{\sigma}^{2} \sigma^{2}~+~
{1\over3}g_{2}\sigma^{3}~+~{1\over4}g_{3}\sigma^{4}
\end{equation}
The scalar potential (2) introduced by Boguta and Bodmer
\cite{BB.77} is essential for appropriate description of
surface properties.  M, m$_{\sigma}$, m$_{\omega}$ and
m$_{\rho}$ are the nucleon-, the $\sigma$-, the $\omega$-
and the $\rho$-meson masses respectively, while
g$_{\sigma}$, g$_{\omega}$, g$_{\rho}$ and e$^2$/4$\pi$ =
1/137 are the corresponding coupling constants for the
mesons and the photon.

The field tensors of the vector mesons and of the
electromagnetic field take the following form:
\begin{equation}
\begin {array}{rl}
\Omega^{\mu\nu} =& \partial^{\mu}\omega^{\nu}-\partial^{\nu}\omega^{\mu}\\
\          \\
\vec R^{\mu\nu} =& \partial^{\mu}\vec\rho^\nu
                  -\partial^{\nu}\vec\rho^\mu\\
\                \\
F^{\mu\nu} =& \partial^{\mu}A^{\nu}-\partial^{\nu}A^{\mu}
\end{array}
\end{equation}
The classical variational principle gives the equations of
motion. In our approach, where the time reversal and charge
conservation is considered, the Dirac equation is written
as:
\begin{equation}
\{ -i{\mbox \boldmath \alpha}{\mbox \boldmath \nabla} +
V({\bf r}) + \beta [ M +S({\bf r}) ] \}
\psi_i~=~\varepsilon_i\psi_i,
\end{equation}
where $V({\bf r})$ represents the vector potential:
\begin{equation}
V({\bf r}) = g_{\omega} \omega_{0}({\bf r}) + g_{\rho}\tau_{3} {\bf {\rho}}
_{0}({\bf r}) + e{1+\tau_{3} \over 2} A_{0}({\bf r}),
\end{equation}
and $S({\bf r})$ is the $scalar$ potential:
\begin{equation}
S({\bf r}) = g_{\sigma} \sigma({\bf r})
\end{equation}
the latter contributes to the effective mass as:
\begin{equation}
M^{\ast}({\bf r}) = M + S({\bf r}).
\end{equation}
The Klein-Gordon equations for the meson fields are
time-independent inhomogeneous equations with the nucleon
densities as sources.
\begin{eqnarray}
\{ -\Delta + m_{\sigma}^{2} \}\sigma({\bf r})
&=& -g_{\sigma}\rho_{s}({\bf r})
-g_{2}\sigma^{2}({\bf r})-g_{3}\sigma^{3}({\bf r})\\
\{ -\Delta + m_{\omega}^{2} \} \omega_{0}({\bf r})
&=& g_{\omega}\rho_{v}({\bf r})\\
\{ -\Delta + m_{\rho}^{2} \}\rho_{0}({\bf r})
&=& g_{\rho} \rho_{3}({\bf r})\\
 -\Delta A_{0}({\bf r}) &=& e\rho_{c}({\bf r})
\end{eqnarray}
The corresponding source densities are
\begin{equation}
\begin{array}{ll}
\rho_{s} =& \sum\limits_{i=1}^{A} \bar\psi_{i}~\psi_{i}\\
\             \\
\rho_{v} =& \sum\limits_{i=1}^{A} \psi^{+}_{i}~\psi_{i}\\
\             \\
\rho_{3} =& \sum\limits_{p=1}^{Z}\psi^{+}_{p}~\psi_{p}~-~
\sum\limits_{n=1}^{N} \psi^{+}_{n}~\psi_{n}\\
\                    \\
\ \rho_{c} =& \sum\limits_{p=1}^{Z} \psi^{+}_{p}~\psi_{p}
\end{array}
\end{equation}
where the sums are taken over the valence nucleons only. It
should also be noted that the present approach neglects the
contributions of negative-energy states ($no-sea$
approximation), i.e. the vacuum is not polarized.

\section{NUMERICAL DETAILS}

The Dirac equation for nucleons is solved using the method
of oscillator expansion as described in Ref. \cite{GRT.90}.
For determination of the basis wavefunctions an axially symmetric
harmonic-oscillator potential with size parameters
\begin{eqnarray}
b_{z}&=&b_{z}(b_{0},\beta_{0})~=~
b_0\exp(\sqrt{5/(16\pi)}\beta_0)\\
b_{\bot}&=&b_{\bot}(b_{0},\beta_{0})~=~
b_0\exp(-\sqrt{5/(64\pi)}\beta_0)
\end{eqnarray}
is employed. The basis is defined in terms of the
oscillator parameter $b_{0}$ and the deformation parameter
$\beta_{0}$. The oscillator parameter $b_0$ is chosen
as $b_0=41 A^{-1/3}$ and the basis deformation $\beta_0$ is
determined for each nucleus in such a way that the
resulting mass quadrupole moment $Q$ of the nucleus is
given by $Q=\sqrt{\frac{16\pi}{5}} \frac{3}{4\pi} A R_0^2\beta_0$ with
$R_0=1.2 A^{1/3}$.

For, most of the nuclei considered here are open-shell
nuclei both in protons and neutrons (except those with
N=82), pairing has been included. We have used the BCS
formalism for the pairing as in our previous works
\cite{GRT.90}.  Constant pairing gaps have been used, which
are taken from the empirical particle separation energies
of neighboring nuclei and in cases where empirical data are
not known, the gaps have been extrapolated according to the
empirical rule $\Delta = 11.2/\sqrt{N(Z)}$ (MeV).
The zero-point energy of an harmonic
oscillator has been used for the center-of-mass energy
correction.  We have neglected the angular momentum and
particle number projection as well as the ground state
correlations induced by coupling to collective
vibrations. It is, however, expected that these additional
corrections would have only small contributions. Here,
therefore, we aim to describe the ground-state properties
of nuclei within the realm of the pure mean field.

The number of oscillator shells taken into account is 12
for both fermionic and bosonic wavefunctions. For
convergence reasons we also considered 14 shells for a
trial. It turned out, however, that the difference in the
results of 12 and 14 shells is negligible.  All the
calculations were hence performed in a basis of 12
harmonic oscillator shells.

The parameter set NL1 has been shown to provide reasonably
good results for nuclei about the line of stability \cite{GRT.90}.
However, because of a very large asymmetry energy 44 MeV of
NL1 as compared to the empirical value, NL1 does not
provide a good description of nuclei away from the
stability line.  This has also the consequence that it
overestimates the neutron-skin thickness \cite{sha92} of
nuclei with large neutron excess.  This problem has been
remedied in the force NL-SH, whereby the $\rho$-meson
coupling constant and thus the asymmetry energy has been
brought very close to the empirical value. In this paper we
have therefore used the force NL-SH. This force has
subsequently been shown to provide excellent results
\cite{SNR.93,SLH.94} for nuclei on both the sides of the
stability line. The parameters of NL-SH have been taken from
Ref. \cite{SNR.93,SLR.93}. The exact values and units of parameters
of the force NL-SH are:

\par\noindent
M = 939.0 MeV; $m_\sigma$ = 526.059 MeV; $m_\omega$ = 783.0 MeV;
$m_\rho$ = 763.0 MeV;
\par\noindent
$g_\sigma$ = 10.444; $g_\omega$ = 12.945; $g_\rho$ = 4.383;
$g_2$ = $-$6.9099 fm$^{-1}$ ; $g_3$ = $-$15.8337.

Charge densities and the corresponding charge radii are
obtained by folding the proton point densities with the
proton form factor of the Gaussian shape. This leads, in a
rather good approximation, to the formula
\begin{equation}
r_c~=~\sqrt{r_p^2 + 0.64},
\end{equation}
where the rms-radius of the proton has been taken to be 0.8
fm.

The quadrupole moments for protons and neutron are
calculated according to the usual definition \cite{LQ.82}
\begin{eqnarray}
Q_2&=&\langle 2r^2 P_2(\cos\theta)\rangle~=~
\langle 2 z^2 - x^2 - y^2\rangle
\label{beta2} \\
Q_4&=&\langle r^4 Y_{40}(\theta)\rangle~=~
\sqrt{\frac{9}{4\pi}}\frac{1}{8}
\langle 8 z^4 - 24 z^2 (x^2+y^2) + 3 (x^2+y^2)^2\rangle
\label{beta4}
\end{eqnarray}

The quadrupole deformation parameter $\beta_2$ and the
hexadecapole deformation parameter $\beta_4$ are obtained
in such a way, that sharp edged densities with this
deformations have the same multipole moments, as discusses
in the appendix of Ref. \cite{LQ.82}.

\section {\sf RESULTS AND DISCUSSION}

We have performed relativistic Hartree calculations for six
isotopic chains in the rare-earth region. Isotopes of Nd
(Z=60), Sm (Z=62),  Gd (Z=64), Dy (Z=66), Er (Z=68) and Yb
(Z=70) have been considered. For, many nuclei in these
chains are known to be well deformed and several shape
transitions along these isotopic chains are expected, we
have solved the RMF equations for an axially deformed
configuration both for prolate as well oblate shape.
However, we present the results only for the lowest energy
shape. Since we do not include triaxial degrees of freedom
we cannot decide whether the second minimum is a local
minimum or a saddle point. On the other hand, empirically
only absolute values of the quadrupole deformations are
known. Some microscopic calculations using the Gogny force
and Skyrme force SIII \cite{QF.78} do predict the
shapes of a few nuclei.  However, extensive microscopic
predictions on the exact shapes are lacking. In this
context, the empirical isotope shifts which are sensitive
to the size and the sign of deformation serve to
reveal the character (signature) of the shapes (Ref.
\cite{Ott.89}).

Calculations for Nd isotopes include mass numbers A=130 up
to A=162, for Sm isotopes A=134 up to A=164, for Dy
isotopes A=142 up to A=168, while the Er and Yb isotopes
cover the region from A=142 to A=170 and A=154 to A=184,
respectively. On the chains of Nd, Sm, Dy, Er and Yb
several precision measurements \cite{Ott.89} on the isotope
shifts are available. We have also performed calculations
for Gd nuclei in the region 136$\leq$A$\leq$172. We compare
our results on this chain with some experimental data
derived from Ref. \cite{ABD.88}, where measurements on
mean-square charge radii of some Gd nuclei have been
performed.

\subsection {\sf Binding Energies}

In Tables 1-6 we present the total binding energies of the
isotopes considered in this work together with the
predictions of the mass formulae Finite Range Droplet Model
(FRDM) \cite{MNM.95} and Extended Thomas-Fermi and
Strutinsky Integral (ETF-SI) \cite{APD.92}. The last column
of the tables shows the available empirical values from the
recently published atomic mass evaluation
\cite{AW.93}. The calculated binding energies in the RMF
theory are in good agreement with the empirical data for
all the isotopic chains and thus the binding energies of a
large number of nuclei have been reproduced by the RMF
theory.  For only a few nuclei full agreement is not
obtained. However the deviations of the RMF values
from the empirical ones is at the most 0.3\%. Our results
are also in overall good agreement with the values of both
the mass models which stretch over all the isotopic
chains and the mass ranges considered here. Only for a few
neutron-deficient Er nuclei is the disagreement with the
models slightly higher than for other isotopes, and is
about 0.4\%.  The agreement of the RMF theory with the
empirical data and with the mass models is here noteworthy,
especially as non-linear the RMF theory entails only 6
parameters fitted to six spherical nuclei at the stability
line. It is worth noting that in contrast both the mass
models FRDM and ETF-SI have been fitted exhaustively to
varying number of parameters to include empirical data of
over one thousand known nuclei. It is, therefore, gratifying
that the RMF theory with only 6 parameters is able to reproduce
the general trend of the absolute binding energies within an
$rms$ deviation of about 2 MeV.

We show in figures 1-2 the neutron and proton single-particle
spectra near the Fermi energy (indicated by dashed line) for the
nucleus $^{166}$Er as a characteristic case. A comparison
is made with the single-particle spectra taken from from the
density-dependent Skyrme forces SIII and SkM \cite{BQB.82}.
The single-particle spectrum from the Modified Harmonic
Oscillator \cite{GLN.67} (MHO) is also shown. The levels are
labeled using the standard notation $\Omega^{\pi}$. The
nucleus $^{166}$Er is highly deformed in its ground state
with $\beta_2 = 0.34$. A comparison of single proton
spectra from the RMF and Skyrme theories shows that the RMF
parameter set NL-SH has larger gaps between levels. The
Skyrme force SkM also shows gaps which are slightly smaller
than those in NL-SH. The level density of spectra in NL-SH
and SkM are similar. In comparison, SIII shows more evenly
distributed single particle levels than both NL-SH and SkM. On the
other hand, the MHO spectrum is denser than both the RMF
and Skyrme forces. It is interesting to note that though
both NL-SH and SIII have a larger compressibility modulus
of about $K = 355$ MeV, this fact does not reflect itself
in the single-particle spectra.

The neutron single-particle spectra shown in figure 2 also
look similar to those of protons except that SkM shows
lesser gaps for neutrons than for protons. The NL-SH
spectrum has clearly higher gaps for the neutrons than the
corresponding spectrum of SkM. On the other hand, the SIII
spectrum is slightly denser than both NL-SH and SkM. The
MHO spectrum is, in this case, much denser than both the
RMF and Skyrme spectra. This can be understood by the
effective mass $m^*/m = 1$ in MHO. The Fermi energy in all the
cases is very close to each other. The difference in the
spin-orbit interaction in the RMF theory and the Skyrme
approach should lead to differences in details of the
single-particle spectra and hence would
also reflect in the magnitude of the nuclear radii.

\subsection{\sf Radii and Isotope shifts}

The nuclear radii obtained in the RMF theory are shown in
figures 3-8.  Here the charge and neutron radii obtained for Nd,
Sm, Gd, Dy, Er and Yb isotopes, respectively, have been
provided. The neutron radii show an increasing trend with
neutron number for all the isotopic chains.  However, for
Er nuclei it shows a clear kink about the magic neutron
number N=82. A much weaker kink about N=82 is also visible
in the neutron radii of Nd, Sm and Dy chains. This is an
indication of the shell effects in these nuclei. Such shell
effects leading to a kink have been observed in the
empirical isotope shifts of Pb nuclei.

The neutron radii for Gd isotopes (Fig. 5) show an
exceptional behaviour in the sense that for the lighter Gd
nuclei and in particular 4-6 neutrons below the closed
neutron shell, nuclei seem to acquire a neutron radius in
unison with the corresponding charge radii, i.e. the
lighter isotopes have a larger size than those of the
heavier magic-neutron counterpart.  This feature has been
symptomatic of the empirical charge radii of Sr and Kr
isotopic chains, the effect which has been described
successfully in the RMF theory. This behaviour of the
neutron radii of Gd nuclei can be put in perspective with
the above feature assuming Z=64 shows some magicity.

The RMF charge radii have been used to obtain isotopes
shifts $r^{2}_{c}(A)-r^{2}_{c}$(ref) for all the isotopic
chains, where $r_c$(ref)  denotes $rms$ charge radius of
the respective reference nuclei. The semi-magic nuclei
$^{142}$Nd, $^{144}$Sm, $^{146}$Gd, $^{148}$Dy, $^{150}$Er
having closed neutron-shell with N=82 have been used as
reference points. Only for the isotopic chain of Yb nuclei,
the nucleus $^{168}$Yb which is without magic neutron
number, has been used as a reference. This has been done
with a view to facilitate the comparison of our predictions
with the empirical data as done by Otten \cite{Ott.89}. In
figures 9-14 we present our results for the isotopic
shifts. The empirical data has also been shown wherever
available.

The experimental isotope shifts for Nd, Sm, and Dy chains
show a value close to zero or slightly positive for nuclei
below N=82, and a pronounced kink in empirical curves can
be seen. This tendency is very much similar to those of the
empirical isotope shifts of Sr and Kr chains \cite{LS.95}. For
the Er and Yb chain, data below N=82 is not available. It can be
seen from figures 9, 10, and 12 that the empirical data on
the Nd, Sm and Dy isotopes on both the sides of the neutron
closed-shell can be very well reproduced in the RMF theory
using the parameter set NL-SH. The kink in the isotope
shifts of our calculations appears for all nuclei except
for Yb where we have not included the semi-magic nucleus
due to the unavailability of empirical data. The kink and
the structure about and below the semi-magic nuclei N=82
stems primarily from the onset of deformation which tend to
overweigh the $rms$ charge radii of the lighter nuclei with
respect to the corresponding magic neutron nuclei. The RMF
theory describes this behaviour very well and predicts a
similar behaviour also for nuclei for which data do not
exist.

For Gd nuclei (Fig. 11) the isotope shifts for the very
light isotopes vary slightly differently than the other
chains in the RMF theory. The isotopic shift for the
nucleus $^{144}$Gd (N=80) is a little negative and then for
even lighter isotopes, the isotopic shifts increase with a
decrease in neutron number down to A=140. This implies that
Gd nucleus with A=144 has the smallest $rms$ charge
distribution and that the nucleus $^{140}$Gd has a much
bigger $rms$ charge radius than the semi-magic nucleus
$^{146}$Gd. This effect is clearly visible in Fig. 5. It
can partly be attributed to a large and abrupt shape
transition from prolate to oblate and again back to prolate
in the region of A=140-144, a behaviour which will become
clearly apparent in Fig. 19.

The experimental isotopic shifts on Gd nuclei have been
measured by a Russian group \cite{ABD.88}. The measurements
pertain to the mean-square charge radii of several Gd
isotopes with neutron numbers above N=82. We have derived
the isotope shifts from this work, which have been shown in
Fig. 11 for comparison. The empirical isotopic shifts
coincide practically with the predictions of the RMF theory.

For the Yb chain, we have shown the results by taking
$^{168}$Yb as a reference nucleus. We observe that the RMF
values lie close to the empirical values on both the sides
of the reference nucleus. An overall view of the isotope
shifts of all the chains except Yb shows that the behaviour
of the charge radii and thus of the isotope shifts is
consistently similar. It can be noted that the RMF theory
is able to reproduce the available data very well.

Changes in charge radii with respect to neighboring
even-even nuclei can be best reflected by the so-called
Brix-Kopferman plot. In Fig. 15 we show the differential
changes of the mean-square charge radii of the neighboring
isotopes, $\delta <r^{2}>^{N-2,N}$, in the Brix-Kopfermann
diagram \cite{BK.49,BK.58}) obtained from the RMF theory. A
strong peak at N=90 can be seen clearly in this quantity
for the isotopic chains of Nd, Sm and Gd. This differential
change decreases considerably above N=90 and then flattens.
The peak at N=90 corresponds to a sudden onset of strong
static deformation for nuclei below N=90 in these isotopic
chains. A small value of the differential change in the
mean-square charge radius above N=94 depicts a saturation in the
value of the quadrupole deformations of nuclei. This effect
will become clear from the magnitude of deformations in the
next figures. It is also to be seen in Fig. 15 that this
peak gradually diminishes going from Nd to Yb.  An
additional small peak also appears at N=86 for some nuclei
such as Gd and Dy.  For the sake of comparison with the
experimental curve, we show in Fig. 16 the Brix-Kopferman
plot derived from the empirical isotope shifts
\cite{Ott.89}. The empirical Brix-Kopferman diagram shows a
structure similar to that obtained in the RMF theory.  The
main peak appears at N=90 as predicted in the RMF theory.
Thus, the RMF theory agrees with the empirical data very well.
This fact has already been seen in the isotope shifts in
figures 9-14. However, a secondary peak in the RMF curve
appearing at N=86 is not to be seen in the empirical data.
Instead a small peak appears at N=84 in the empirical
curve. This discrepancy in the plots about N=84-86 can
be attributed to a complex and slightly different evolution
of the shapes and the magnitudes of deformation on adding
neutrons to the closed-shell (N=82) nuclei.

\subsection {\sf Deformations Properties}

\subsubsection{\sf Quadrupole deformation}

The deformations and shapes of nuclei play a crucial role
in defining the properties such as nuclear sizes and
isotope shifts.  In the RMF theory we have obtained the
quadrupole and hexadecapole moments of nuclei from the solution
of deformed RMF equations. The resulting quadrupole and hexadecapole
deformation parameters $\beta_2$ and $\beta_4$ are calculated
using the method of Ref. \cite{LQ.82} as discussed at the end
of Section III.

The $\beta_2$ values are shown in Figs. 17-22 for all the
isotopic chains considered in this work. Since the mass
formulae FRDM and ETF-SI provide the deformations obtained
from their exhaustive fits, we supplement the figures with
the predictions of these models for the sake of comparison.
As the empirical values of $\beta_2$ obtained from BE(2)
values do not contain the sign of the deformation, we do
not show the empirical values in these figures. However, the
empirical values will be given in the tables below.

The $\beta_2$ values obtained in the RMF theory manifest an
interesting change of shapes of nuclei below and above the
magic neutron number N=82.  Interestingly, we find that for
most of the chains, there is an excellent agreement of the RMF
predictions with the mass models.  In greater number of
cases the RMF values are much closer to the FRDM than the
ETF-SI both in the variation (including shape transitions)
as well as in the magnitude. For nuclei such as Nd and Sm
(Figs. 17 and 18), the ETF-SI model predicts a slightly
different variation of deformation with mass number than
that predicted by the RMF theory and the FRDM.

For most of the chains, there is an onset of prolate
deformation above N=82. All the nuclei with N=82 are
spherical as expected. However, a successive addition of
neutrons to the magic neutron core leads to a gradual
evolution of prolate shape in the RMF theory.  The nucleus
$^{148}$Gd is an exception where the RMF theory predicts a
slightly oblate shape. For neutron numbers higher than
N=82, the prolate deformation increases and then saturates
at a value close to $\beta_2 = 0.30-0.35$ in most of the
cases. In comparison, both the FRDM and ETF-SI also predict
an increasing prolate deformation for nuclei above N=82.
The ETF-SI predicts an early onset of the prolate
deformation as compared with NL-SH and FRDM for most of the
chains considered here. The quadrupole deformation values
in the ETF-SI for nuclei immediately closer to N=82 are
higher than both NL-SH and the FRDM. This behaviour can be
seen in Figs. 18-21.  In general, there is a striking
similarity in the predictions of the RMF theory and the
FRDM and ETF-SI results for nuclei above N=82.

For nuclei below N=82, all the isotopic chains (except Yb
which does not encompass N=82) exhibit an almost identical
behaviour in the quadrupole deformation. Nuclei in all the
chains (Figs. 17-21) undergo a spherical to oblate shape
transition below the neutron magic number except for Gd (Fig. 19)
and Dy (Fig. 20) nuclei, where nuclei just below magic
number (N=80) acquire a prolate shape in the RMF theory.
The sharp transition to the oblate shape in the five chains
is taking place at N=78 in the RMF theory.  Only in the
case of Er this transition occurs smoothly from a spherical
shape via a slightly less oblate shape for $^{148}$Er
(N=80). All nuclei below N=78 suddenly acquire a prolate
shape the magnitude of which increases for Nd and Sm
isotopes with a decrease in the neutron number.  It is
worth noting that the oblate shape at N=78 for all chains,
the prolate shape at N=80 for Gd and Dy and the shape
transition from oblate to prolate in going to lower neutron
numbers below N=78 are the aspects which are meticulously
consistent in the RMF theory and in the FRDM for all the above
five chains.  The ETF-SI does show these features for Gd,
Dy and Er (Figs.  19-21) and is consistent both with NL-SH
as well as FRDM in these complex series of transitions.
However, the ETF-SI tends to predict a slightly higher prolate
deformation for the above nuclei than that predicted by
NL-SH and FRDM.  For the isotopic chains of Nd and Sm
(Figs. 17-18) and for nuclei below N=82, the ETF-SI values
deviate strongly from the predictions of NL-SH and FRDM.
For Nd (Z=60) isotopes ETF-SI predicts an increasingly
prolate shape for nuclei below and including N=82, without
giving a spherical shape to the neutron magic nucleus
$^{142}$Nd. The NL-SH and FRDM provide a spherical shape to
nuclei two neutron numbers below and above N=82 for Nd and Sm.
Thus, in the five isotopic chains (excluding Yb) the RMF theory
as well as FRDM predict  a complex series of
prolate-oblate-spherical-prolate shape
transitions with increasing neutron number.

The chain of Yb (Z=70) isotopes does not include a magic
neutron number. The quadrupole deformations for nuclei
above N=84 (Fig. 22) show an increasingly prolate shape
with an increase in neutron number in the RMF theory as
well as in FRDM and ETF-SI.  The $\beta_2$ value saturates
above N=98 in all the three approaches. The magnitude of
the $\beta_2$ value in NL-SH is in some cases closer to
FRDM and in others closer to ETF-SI. The general trend for
Yb isotopes is similar in all the three approaches.

The numerical values of the quadrupole deformations in the RMF
theory are given in Tables 7-12. The $\beta_2$ values from
the FRDM and ETF-SI are also provided in the tables for the
sake of comparison of magnitude and signature. The absolute
values of the empirical $\beta_2$ obtained from the
compilation of Raman et al. \cite{RMM.87} are shown in the
last columns. It must be stressed that these values do not
indicate the sign of the quadrupole moment. For the
nuclei where the experimental value of the deformation has
been deduced, the $\beta_2$ values from NL-SH correspond
closely to the empirical ones in magnitude. For some
specific nuclei such as $^{152}$Sm and $^{154}$Sm, the
$\beta_2$ values obtained by fitting the differential
cross-sections in electron scattering
experiments \cite{CBH.76} have been given in the last
column of Table 8 in the parentheses. The RMF values show a
significant agreement with these empirical values. Thus,
the RMF theory describes all the available data on the
deformation of the rare-earth nuclei successfully. The detailed
behaviour of shapes and shape transitions has already been
alluded to above.

A comparison of the NL-SH values and the empirical
quadrupole deformation has been facilitated in Fig. 23,
where we show $\beta_2^2$ for all the isotopic chains. The
figure shows an overall consistency of the theoretical
values with the empirical ones. However, in presenting the
squares of $\beta_2$ any slight discrepancy is likely to be
amplified. Taking this fact into account, there is a broad
agreement between the RMF deformations and the empirical
values. This is corroborated by our Tables 7-12.

In order to judge the comparative behaviour of various
isotopic chains towards the quadrupole deformation, we show
the loci of $<\beta^{2}_{2}>$ in Fig. 24, as a function the
neutron number.  The minima at the magic number N=82 are
clearly visible. On both the sides of this number, a parabola
like behaviour is to be seen.  Nuclei such as Nd, Sm and Gd
exhibit strong shape transition and highly deformed shapes
above N=88. They reach a saturation at about N=102. For the
other chains such as Dy, Er and Yb, this trend is much more
gradual. For Yb isotopes the maximum deformation is a
little less than those of its neighboring chains.

\subsubsection{\sf Hexadecapole deformations}

The hexadecapole deformation has usually been inferred by
fitting cross-sections in inelastic scattering experiments
\cite{HGH.68} and from experiments with Coulomb excitation
\cite{WWE.74}.  Empirical data on the hexadecapole
deformations of nuclei is available only scantily. On the
other hand, most of the mass formulae such as FRDM and
ETF-SI employ a hexadecapole degree of freedom in the
minimization in their exhaustive and global fits.  The FRDM
and the ETF-SI thus predict the hexadecapole deformation
$\beta_4$ along with the quadrupole deformation $\beta_2$
which has been discussed above.

First, we show the hexadecapole deformation $\beta_4$ obtained
with NL-SH for all the isotopic chains in Tables 13-18. Predictions
of the FRDM and ETF-SI are also shown for comparison.
For all the isotopic chains except Yb (which does not
include a magic neutron number), nuclei below N=82 exhibit
a negative value of $\beta_4$. Comparing this with the sign
of the $\beta_2$, this is in contrast with the
overwhelmingly positive $\beta_2$ values for neutron numbers
in this region, except for the oblate shape at N=78. The
$\beta_4$ values then edge to zero at and around the magic
numbers where nuclei are expected to be spherical. This is
consistent with the corresponding quadrupole deformation
$\beta_2$ which vanishes about N=82. The FRDM also
predicts negative $\beta_4$ values below N=82.  This is
again in consistency with the RMF predictions.  The ETF-SI, on
the other hand, is consistent with RMF and FRDM for Gd, Dy
and Er isotopes. In contrast to RMF and FRDM, ETF-SI
gives positive $\beta_4$ values for the Nd and Sm nuclei
below N=82. A comparison of the theoretical predictions
with the empirical hexadecapole moments will be made below,
wherever experimental data are available.

For nuclei above N=82, the RMF theory predicts a positive
$\beta_4$ for all the chains except Yb (Table 18). The
corresponding quadrupole deformation for all the chains
including Yb have a prolate shape.  Only for the Yb chain,
the $\beta_4$ values are negative in clear contrast with
all other chains, though all the Yb isotopes above N=82
show a prolate quadrupole shape. Again, the predictions of
the RMF theory on the sign of the hexadecapole deformation are
consistent with those of the FRDM and ETF-SI for all the
chains including that of Yb.

In Tables 19-24 we show our predictions on the charge
hexadecapole moment calculated with the force NL-SH. The
experimental data from a very recent compilation by
L\"obner \cite{loeb95} are also shown where available. It
may be noted that the same convention for the empirical
data is used. In some cases more than one values is
provided as having been obtained using different probes. In
many cases, where more than one empirical value for a
nucleus is given, there is a unanimity in the values. In
the others the empirical values indicate a rather broad
differences and sometimes the error bars are too large to
reach a conclusion on the value. However, in general, there
is a broad and good agreement of the RMF hexadecapole
moments with the empirical data notwithstanding the
inherent uncertainties which some data possess.

\section{\sf Summary and Conclusions}

The relativistic mean-field theory with the non-linear
$\sigma\omega$ model has been employed to investigate the
ground-state properties of several chains of rare-earth
nuclei. The calculations have been performed for deformed
axially symmetric configurations in the relativistic mean-field
approximation for the even-even nuclei of the isotopic
chains of Nd, Sm, Gd, Dy, Er and Yb. Binding energies,
isotopic shifts and deformation properties have been
obtained using the parameter set NL-SH. The results of these
calculations have been compared with the empirical data
available on the binding energies, isotopic shifts, and
quadrupole and hexadecapole deformations.

The RMF theory describes the binding energies of nuclei
over a large range of proton and neutron numbers very well.
The RMF binding energies are also in good agreement with
the predictions of the extensive mass formulae FRDM and
ETF-SI. The empirical isotopic shifts of all the isotopic
chains have been described successfully. It is noteworthy
that the behaviour of the isotopic shifts in the deformed
rare-earth nuclei is reminiscent of those of the Sr and Kr
isotopic chains, the data which have earlier been
reproduced \cite{LS.95} well only within the RMF theory.

The quadrupole deformations predicted by the RMF theory have
been found to be in very good agreement with the empirical
data. The magnitude of the quadrupole deformation $\beta_2$
in the RMF theory shows a good agreement also with those of
FRDM and ETF-SI. In addition, the RMF theory predicts a complex
web of shape transitions, which are observed to be similar
for most of the isotopic chains. The rare-earth nuclei in
this region exhibit shape transitions
prolate-oblate-spherical-prolate for all the chains (except
Yb) with an increasing neutron number. The shape transition
from prolate to oblate at N=78 is spectacular as nuclei on
both the sides of N=78 assume deformations in the direction
of positive $\beta_2$. The complex series of shape
transitions both below and above the neutron magic number
N=82 are in astonishing agreement with the FRDM. With some
exceptions, the RMF results also show good agreement with
ETF-SI.

The hexadecapole moments and $\beta_4$ values obtained in
the RMF theory show a good comparison with the
corresponding empirical values wherever available. The RMF
theory provides a negative $\beta_4$ values for nuclei
below the neutron magic number N=82.  In contrast, the
$\beta_4$ values for nuclei above N=82 are positive for all
the chains. For the case of Yb chain only, nuclei above
N=100 exhibit negative $\beta_4$ values. Comparing with the
mass models, a very good agreement of the RMF predictions
on the $\beta_4$ values has been observed with the
predictions of the FRDM for all the isotopic chains. Again,
with a few exceptions only, the RMF values are in accord
with the ETF-SI predictions.

It may be reckoned \cite{LS.95} that the RMF theory with the
force NL-SH has been successful in providing a good
description of the anomalous isotopic shifts of Sr and Kr
nuclei which undergo a series of complex shape transitions
as also is the case with the rare-earth nuclei. The present
calculations on the rare-earth nuclei in conjunction with
the results of Sr and Kr nuclei demonstrate the ability of
the RMF theory with the force NL-SH to describe a broad range
of nuclear data encompassing a large range of nuclei and
isospins.

In retrospect, calculations using non-relativistic Skyrme
and Gogny forces were carried out for deformed nuclei by
several groups. Extensive calculations of entire chains of
isotopes with these approaches are not available.  On the
other hand, some of these approaches using the Skyrme force
SIII and the Gogny force D1 have been able to reproduce
results on deformation properties such as quadrupole and
hexadecapole moments and on binding energies for a set of
selected nuclei. However, these approaches have not been
able to provide an adequate description of the anomalous
nature of the isotopic shifts in deformed rare-earth
nuclei \cite{Ott.89}. Also, it is not clear how these
approaches will respond to the properties over a large
range of isospin.

A correct description of the isotopic shifts depends
crucially upon the deformations assumed by nuclei. The
ability of the RMF theory in describing the isotopic shifts
of rare-earth nuclei as well as those of Sr and Kr nuclei
stems from the successful description of the deformations.
In this context, the shell effects play a major role in
determining the potential energy landscape and consequently
the deformation of the ground state.  It has been pointed
out \cite{SLK.95} that shell effects in the RMF theory are
different from those in the non-relativistic approaches.
Thus, the appropriate shell effects render a unified and
comprehensive description of several aspects of
the ground-state properties of nuclei.

\section{\sf Acknowledgment}
One of the authors (G.A.L) acknowledges support by the
European Union under the contract HCM-EG/ERB CHBICT-930651.
Partial support from the Bundesministerium f\"ur Forschung
und Technologie under the project 06TM734(6) is acknowledged.
We thank Prof. K.E.G. L\"obner for supplying us his
compilation on hexadecapole moments prior to its
publication.


\newpage
\leftline{\bf Figure Captions}
\bigskip
\noindent
{\bf Fig. 1} The neutron single-particle (s.p.) spectrum for $^{166}$Er
near Fermi energy obtained in the RMF theory using the force NL-SH.
The spectra from the Skyrme forces SIII and SkM are shown for comparison.
The corresponding spectrum using the Modified Harmonic Oscillator (MHO)
is also given.
\bigskip

\noindent
{\bf Fig. 2} The proton s.p. energies for $^{166}$Er. For details,
see the caption of Fig. 1.
\bigskip

\noindent
{\bf Fig. 3.} The charge and neutron radii of Nd isotopes in the RMF
theory. For details, see text.
\bigskip

\noindent
{\bf Fig. 4}  The charge and neutron radii of Sm isotopes in the RMF
theory. For details, see text.
\bigskip

\noindent
{\bf Fig. 5}  The charge and neutron radii of Gd isotopes in the RMF
theory. For details, see text.
\bigskip

\noindent
{\bf Fig. 6}  The charge and neutron radii of Dy isotopes in the RMF
theory. For details, see text.
\bigskip

\noindent
{\bf Fig. 7}  The charge and neutron radii of Er isotopes in the RMF
theory. For details, see text.
\bigskip

\noindent
{\bf Fig. 8}  The charge and neutron radii of Yb isotopes in the RMF
theory. For details, see text.
\bigskip

\noindent
{\bf Fig. 9} The isotope shifts for Nd nuclei obtained in the RMF
theory with the force NL-SH. The empirical data {\protect\cite{Ott.89}}
are also shown for comparison. A kink about the N=82 nucleus can been
clearly.
\bigskip

\noindent
{\bf Fig. 10} The same as in Fig. 9, for Sm nuclei.
\bigskip

\noindent
{\bf Fig. 11} The same as in Fig. 9. However, the empirical
data has been obtained from Ref. \cite{ABD.88}.
\bigskip

\noindent
{\bf Fig. 12} The same as in Fig. 9, for Dy nuclei.
\bigskip

\noindent
{\bf Fig. 13} The same as in Fig. 9, for Er nuclei.
\bigskip

\noindent
{\bf Fig. 14} The isotope shifts for Yb nuclei. The nucleus
$^{168}$Yb has been used as a reference point.
\bigskip

\noindent
{\bf Fig. 15} The Brix-Kopfermann plot obtained in the RMF
theory with the force NL-SH.
\bigskip

\noindent
{\bf Fig. 16} The Brix-Kopferman plot using the experimental
isotope shifts.
\bigskip

\noindent
{\bf Fig. 17} The quadrupole deformation $\beta_2$ for Nd isotopes
with NL-SH. The predictions of the mass models FRDM and ETF-SI are
also shown for comparison. See text for details.
\bigskip

\noindent
{\bf Fig. 18} The same as in Fig. 17, for Sm isotopes.
\bigskip

\noindent
{\bf Fig. 19} The same as in Fig. 17, for Gd isotopes.
\bigskip

\noindent
{\bf Fig. 20} The same as in Fig. 17, for Dy isotopes.
\bigskip

\noindent
{\bf Fig. 21} The same as in Fig. 17, for Er isotopes.
\bigskip

\noindent
{\bf Fig. 22} The same as in Fig. 17, for Yb isotopes.
\bigskip

\noindent
{\bf Fig. 23} $\beta_2^2$ for various rare-earth nuclei
in the RMF theory (open circles). The available experimental
values (solid circles) \cite{RMM.87} from BE(2) measurements are
also shown for comparison. For details see text.
\bigskip

\noindent
{\bf Fig. 24} Loci of $\beta_2^2$ obtained in the RMF theory,
for various isotopic chains of rare-earth nuclei. A parabolic
behaviour about the magic neutron number is observed.
\newpage

\newpage
\noindent\begin{table}
\begin{center}
\caption{\sf The binding energies (in MeV) for Nd isotopes obtained
with the force NL-SH. The predictions from the mass models FRDM
and ETF-SI are also shown for comparison. The empirical
values (expt.) are shown in the last column.}
\bigskip
\begin{tabular}{ll c c c c c l}
\hline\hline
&A& NL-SH& FRDM& ETF-SI&  expt.&\\
\hline
&130&1070.39&1069.10&1068.82&1068.67& \\
&132&1091.64&1089.97&1090.16&1090.09& \\
&134&1112.01&1109.75&1109.86&1110.38& \\
&136&1131.39&1129.09&1128.92&1129.92& \\
&138&1151.42&1147.92&1148.37&1148.94& \\
&140&1172.87&1167.15&1167.27&1167.52& \\
&142&1190.19&1185.67&1185.85&1185.15& \\
&144&1203.39&1199.16&1199.28&1199.09& \\
&146&1214.46&1212.00&1212.52&1212.41& \\
&148&1226.05&1225.33&1225.56&1225.03& \\
&150&1238.39&1238.46&1238.35&1237.45& \\
&152&1251.07&1250.93&1250.48&1250.06& \\
&154&1262.03&1262.45&1261.72&1261.66& \\
&156&1272.74&1273.01&1272.33&1272.29& \\
&158&1282.36&1282.88&1282.21&  -    & \\
&160&1292.42&1292.14&1291.24&  -    & \\
&162&1300.43&1300.81&1299.33&  -    & \\
\hline\hline
\end{tabular}
\end{center}
\end{table}
\newpage
\noindent\begin{table}
\begin{center}
\caption{\sf The binding energy (in MeV) of Sm isotopes.
For details refer to the caption of Table 1}
\bigskip
\begin{tabular}{ll c c c c c l}
\hline\hline
& A & NL-SH & FRDM & ETF-SI & expt.&\\
\hline
&134& 1097.82&1094.65  &1095.16 & 1094.51 &\\
&136& 1118.91&1115.72 & 1116.09&1115.98   &\\
&138& 1138.85 & 1136.17&1135.89&1136.56 &\\
&140& 1159.90&1156.00 & 1156.09&1156.47 &\\
&142& 1180.37&1176.29& 1176.31&1176.61 &\\
&144& 1200.91&1196.09&1196.27&1195.74 &\\
&146& 1213.98&1210.82&1210.91&1210.91 &\\
&148& 1227.99&1225.00&1225.67&1225.40& \\
&150& 1241.40&1239.67&1240.23&1239.25& \\
&152& 1255.75&1254.11&1254.22&1253.12& \\
&154& 1269.60&1267.87&1267.83&1266.94& \\
&156& 1281.94&1280.72&1280.39&1279.99& \\
&158& 1294.40&1292.59&1291.91&1292.03& \\
&160& 1305.34&1303.86&1303.35&1303.19& \\
&162& 1315.35&1314.42&1313.48& -&\\
&164& 1324.63&1324.25&1322.98& -&\\
\hline\hline
\end{tabular}
\end{center}
\end{table}
\newpage

\noindent\begin{table}
\begin{center}
\caption{\sf The binding energies for Gd isotopes. See the caption of
Table 1 for details.}
\bigskip
\begin{tabular}{ll c c c c c l}
\hline\hline
&A& NL-SH& FRDM& ETF-SI&  expt.&\\
\hline
&138&1123.32&1119.77&1119.82&1119.69&\\
&140&1144.41&1141.48&1141.12&1141.70&\\
&142&1165.99&1162.27&1162.30&1163.48&\\
&144&1183.84&1183.54&1183.37&1183.55&\\
&146&1207.42&1204.45&1204.73&1204.44&\\
&148&1222.86&1220.77&1220.50&1220.76&\\
&150&1239.31&1236.07&1236.74&1236.40&\\
&152&1254.51&1251.88&1252.68&1251.49&\\
&154&1269.44&1267.38&1267.95&1266.63&\\
&156&1284.27&1282.24&1282.56&1281.60&\\
&158&1298.31&1296.23&1296.57&1295.90&\\
&160&1311.72&1309.47&1309.38&1309.29&\\
&162&1323.80&1322.00&1321.50&1321.77&\\
&164&1335.02&1333.81&1333.52&1333.37&\\
&166&1345.94&1344.81&1344.09&  -    &\\
\hline\hline
\end{tabular}
\end{center}
\end{table}
\newpage
\noindent\begin{table}
\begin{center}
\caption{\sf The binding energies for Dy isotopes. See the caption
of Table 1 for details.}
\bigskip
\begin{tabular}{ll c c c c c l}
\hline\hline
&A& NL-SH&  FRDM& ETF-SI& expt.&\\
\hline
&142&1148.15&1144.78&1144.30&1144.42&\\
&144&1170.83&1166.67&1166.91&1167.39&\\
&146&1191.24&1189.00&1188.62&1189.64&\\
&148&1213.60&1211.24&1211.22&1210.83&\\
&150&1230.70&1228.50&1228.20&1228.40&\\
&152&1248.99&1245.14&1245.67&1245.33&\\
&154&1265.66&1262.08&1262.88&1261.75&\\
&156&1281.93&1278.61&1279.40&1278.04&\\
&158&1297.73&1294.55&1295.24&1294.06&\\
&160&1312.94&1309.69&1310.30&1309.47&\\
&162&1327.17&1324.17&1324.56&1324.12&\\
&164&1340.21&1337.98&1337.98&1338.05&\\
&166&1352.99&1350.99&1350.60&1350.81&\\
&168&1365.63&1263.26&1363.11&1362.82&\\
\hline\hline
\end{tabular}
\end{center}
\end{table}
\newpage
\noindent\begin{table}
\begin{center}
\bigskip
\caption{\sf The binding energies for Er isotopes. See the caption of
Table 1 for details.}
\bigskip
\begin{tabular}{ll c c c c c l}
\hline\hline
&A& NL-SH& FRDM& ETF-SI&  expt.&\\
\hline
&142&1127.36&1121.81&1121.16&   -   & \\
&144&1150.42&1146.00&1144.91&   -   & \\
&146&1173.84&1169.06&1169.02&1169.97& \\
&148&1197.49&1192.59&1192.31&1193.11& \\
&150&1219.12&1215.78&1215.84&1216.15& \\
&152&1237.40&1234.32&1234.00&1234.19& \\
&154&1257.26&1252.25&1252.54&1252.40& \\
&156&1275.45&1270.15&1270.96&1270.39& \\
&158&1288.95&1287.64&1288.67&1286.47& \\
&160&1307.23&1304.70&1305.68&1304.27& \\
&162&1324.00&1321.00&1321.95&1320.70& \\
&164&1339.53&1336.69&1337.31&1336.45& \\
&166&1353.29&1351.75&1352.17&1351.57& \\
&168&1366.65&1366.03&1366.05&1365.78& \\
&170&1379.93&1379.51&1379.09&1379.78& \\
\hline\hline
\end{tabular}
\end{center}
\end{table}
\newpage
\noindent\begin{table}
\begin{center}
\bigskip
\caption{\sf The binding energies for Yb isotopes. See the caption
of Table 1 for details.}
\bigskip
\begin{tabular}{ll c c c c c l}
\hline\hline
&A& NL-SH& FRDM& ETF-SI&  expt.&\\
\hline
&154&1241.63&1238.36&1238.12&1238.97& \\
&156&1262.73&1257.42&1257.48&1257.67& \\
&158&1281.68&1276.32&1276.87&1276.53& \\
&160&1298.55&1294.80&1295.70&1294.81& \\
&162&1315.78&1312.70&1313.84&1312.64& \\
&164&1333.89&1330.19&1331.26&1329.93& \\
&166&1350.45&1347.03&1347.84&1346.67& \\
&168&1365.36&1363.22&1363.71&1362.79& \\
&170&1379.96&1378.66&1379.05&1378.13& \\
&172&1394.22&1393.31&1393.27&1392.77& \\
&174&1408.23&1407.00&1406.95&1406.60& \\
&176&1420.62&1420.09&1419.76&1419.29& \\
&178&1433.41&1432.78&1431.31&1431.63& \\
&180&1443.76&1443.69&1442.41&   -   & \\
&182&1453.79&1454.33&1452.92&   -   & \\
&184&1462.70&1464.34&1462.63&   -   & \\
\hline\hline
\end{tabular}
\end{center}
\end{table}
\vfill
\noindent\begin{table}
\begin{center}
\bigskip
\caption{\sf The quadrupole deformations $\beta_2$ for Nd isotopes
obtained in the RMF theory using the force NL-SH. The FRDM and ETF-SI
predictions are also shown. The available empirical deformations (expt.)
obtained from BE(2) values are also given in the last column. The
experimental values do not depict the sign of the deformation.}
\bigskip
\begin{tabular}{lll c c c c c  l}
\hline\hline
& A & NL-SH &FRDM&ETF-SI& expt.& \\
\hline
&130&0.322&0.311&0.36&  -  & \\
&132&0.268&0.293&0.37&  -  & \\
&134&0.229&0.218&0.36&  -  & \\
&136&0.182&0.171&0.36&  -  & \\
&138&-0.091&-0.138&0.19& - & \\
&140&0.000&0.000&0.15& -& \\
&142&0.000&0.000&0.08&0.093 &\\
&144&0.000&0.000&0.08&0.131& \\
&146&0.080&0.161&0.14&0.152&\\
&148&0.118&0.206&0.21&0.204&\\
&150&0.264&0.243&0.24&0.289&\\
&152&0.318&0.262&0.29&0.274&\\
&154&0.331&0.270&0.31& -&\\
&156&0.338&0.279&0.31& -&\\
&158&0.345&0.279&0.31& -&\\
&160&0.349&0.290&0.32& -&\\
&162&0.350&0.300&0.32& -&\\
\hline\hline
\end{tabular}
\end{center}
\end{table}
\newpage
\noindent\begin{table}
\begin{center}
\bigskip
\caption{\sf The quadrupole deformations $\beta_2$ for Sm isotopes.
See Table 7 for details. The empirical values given in the
parentheses in the last column are from electron scattering
experiments of Ref.\protect{\cite{CBH.76}} }.
\bigskip
\begin{tabular}{lll c  c c c cl}
\hline\hline
& A & NL-SH  & FRDM& ETF-SI& expt.& \\
\hline
&134&  0.301&  0.312&  0.370& -& \\
&136&  0.263&  0.237&  0.360& -& \\
&138&  0.214&  0.190&  0.270& 0.225& \\
&140& -0.124& -0.148&  0.150& -& \\
&142&  0.0  &  0.00 &  0.09 & -& \\
&144&  0.0  &  0.00 &  0.00 & 0.088& \\
&146&  0.0  &  0.0  &  0.120& -& \\
&148&  0.108&  0.161&   0.200& 0.142& \\
&150&  0.166&  0.206&   0.230& 0.193& \\
&152&  0.261&  0.243&   0.260& 0.306 ($0.287 \pm 0.003$)\cite{CBH.76}& \\
&154&  0.309&  0.243&   0.280& 0.341 ($0.311 \pm 0.003$)\cite{CBH.76}& \\
&156&  0.324&  0.279&   0.300&  -& \\
&158&  0.335&  0.279&   0.310&  -& \\
&160&  0.344&  0.290&  0.310 &  -& \\
&162&  0.349&  0.300&  0.320 &  -&\\
&164&  0.350&  0.320&  0.340 &  -&\\
\hline\hline
\end{tabular}
\end{center}
\end{table}
\newpage
\noindent\begin{table}
\begin{center}
\bigskip
\caption{\sf The quadrupole deformations $\beta_2$ for Gd isotopes.
See Table 7 for details.}
\bigskip
\begin{tabular}{lll c c c c c  l}
\hline\hline
& A & NL-SH &FRDM&ETF-SI& expt.& \\
\hline
&138&0.295&0.256&0.37& -  &\\
&140&0.307&0.210&0.33& -  &\\
&142&-0.158&-0.156&-0.21& -&\\
&144&0.081&0.000&0.06&-  &\\
&146&0.000&0.000&0.0& -& \\
&148&-0.063&0.000&0.12& -&\\
&150&0.140&0.161&0.20& -& \\
&152&0.185&0.207&0.25&0.212&\\
&154&0.264&0.243&0.27&0.310&\\
&156&0.314&0.271&0.30&0.338&\\
&158&0.330&0.271&0.30&0.348&\\
&160&0.343&0.280&0.31&0.353&\\
&162&0.350&0.291&0.32& - &\\
&164&0.356&0.301&0.33& - &\\
&166&0.357&0.303&0.34& - &\\
\hline\hline
\end{tabular}
\end{center}
\end{table}
\newpage
\noindent\begin{table}
\begin{center}
\bigskip
\caption{\sf The quadrupole deformations $\beta_2$ for Dy isotopes. See
Table 7 for details.}
\bigskip
\begin{tabular}{lll c c c c  cl}
\hline\hline
& A & NL-SH & FRDM&ETF-SI& expt.& \\
\hline
&142& 0.250&0.219&0.32   & -   &\\
&144& -0.161&-0.164&-0.21& -   &\\
&146& 0.067& 0.000& 0.06&  -   & \\
&148& 0.000& 0.000& 0.02&  -   & \\
&150& 0.077& 0.000& 0.12&  -   & \\
&152& 0.148& 0.153& 0.21& 0.086& \\
&154& 0.187& 0.207& 0.24& 0.237& \\
&156& 0.236& 0.235& 0.26& 0.293& \\
&158& 0.284& 0.262& 0.29& 0.326& \\
&160& 0.299& 0.272& 0.30& 0.337& \\
&162& 0.320& 0.281& 0.32& 0.341& \\
&164& 0.335& 0.292& 0.33& 0.348& \\
&166& 0.349& 0.293& 0.33&  -   & \\
&168& 0.345& 0.304& 0.33&  -   & \\
\hline\hline
\end{tabular}
\end{center}
\end{table}
\newpage
\noindent\begin{table}
\begin{center}
\bigskip
\caption{\sf The quadrupole deformations $\beta_2$ for Er  isotopes. See
Table 7 for details.}
\bigskip
\begin{tabular}{lll c c c c c  l}
\hline\hline
& A & NL-SH &FRDM&ETF-SI& expt.& \\
\hline
&142&0.282&0.277&0.35&-&\\
&144&0.246&0.220&0.33&-&\\
&146&-0.166&-0.173&-0.21&-&\\
&148&-0.131&-0.156&-0.20&-&\\
&150&0.000&-0.008&0.02&-&\\
&152&0.080&-0.018&0.12&-&\\
&154&0.139&0.143&0.18&-&\\
&156&0.175&0.189&0.24&0.189&\\
&158&0.229&0.216&0.26&0.254&\\
&160&0.266&0.253&0.28&0.303&\\
&162&0.289&0.272&0.31&0.322&\\
&164&0.305&0.273&0.33&0.333&\\
&166&0.319&0.283&0.31&0.342&\\
&168&0.333&0.294&0.33&0.338&\\
&170&0.339&0.296&0.33&0.336&\\
\hline\hline
\end{tabular}
\end{center}
\end{table}
\newpage
\noindent\begin{table}
\begin{center}
\bigskip
\caption{\sf The quadrupole deformations $\beta_2$ for Yb  isotopes. See
Table 7 for details.}
\bigskip
\begin{tabular}{lll c c c c c  l}
\hline\hline
& A & NL-SH &FRDM&ETF-SI& expt.& \\
\hline
&154&0.091&-0.008&0.12& -&\\
&156&0.134&0.125&0.16&-& \\
&158&0.165&0.161&0.21& 0.193&\\
&160&0.206&0.208&0.25&0.222&\\
&162&0.245&0.225&0.28&0.262&\\
&164&0.279&0.264&0.29&0.289&\\
&166&0.302&0.274&0.31&0.312&\\
&168&0.311&0.284&0.31&0.327&\\
&170&0.310&0.295&0.33&0.323&\\
&172&0.308&0.296&0.32&0.330&\\
&174&0.305&0.287&0.33&0.325&\\
&176&0.303&0.278&0.31&0.309&\\
&178&0.296&0.279&0.31& - &\\
&180&0.289&0.279&0.30& - &\\
&182&0.278&0.272&0.29& - &\\
&184&0.232&0.233&0.28& - &\\
\hline\hline
\end{tabular}
\end{center}
\end{table}
\newpage
\noindent\begin{table}
\begin{center}
\bigskip
\caption{\sf The hexadecapole deformations $\beta_4$ for Nd isotopes
obtained in the RMF theory using the force NL-SH. The FRDM and ETF-SI
predictions for $\beta_4$ are also shown.}
\bigskip
\begin{tabular}{lll c c c c   l}
\hline\hline
& A  &       &  $\beta_4$&   &\\
\hline
&  &  NL-SH &FRDM&ETF-SI& \\
\hline
&130&-0.010& 0.002& 0.020&\\
&132&-0.024&-0.002& 0.020&\\
&134&-0.033&-0.023& 0.030&\\
&136&-0.033&-0.030&-0.010&\\
&138&-0.009&-0.031& 0.000&\\
&140& 0.000& 0.000& 0.000&\\
&142& 0.000& 0.000&-0.010&\\
&144& 0.000& 0.000& 0.010&\\
&146& 0.009& 0.068& 0.040&\\
&148& 0.033& 0.083& 0.050&\\
&150& 0.093& 0.107& 0.070&\\
&152& 0.114& 0.128& 0.070&\\
&154& 0.104& 0.114& 0.070&\\
&156& 0.085& 0.098& 0.070&\\
&158& 0.067& 0.082& 0.050&\\
&160& 0.041& 0.069& 0.030&\\
&162& 0.020& 0.048& 0.030&\\
\hline\hline
\end{tabular}
\end{center}
\end{table}
\newpage
\noindent\begin{table}
\begin{center}
\bigskip
\caption{\sf The hexadecapole deformations $\beta_4$ for Sm isotopes.
See Table 13 for details.}
\bigskip
\begin{tabular}{lll c  c c c l}
\hline\hline
& A  &     &  $\beta_4$&    &\\
\hline
&  &  NL-SH  & FRDM& ETF-SI&  \\
\hline
&134&-0.019&-0.006& 0.020&\\
&136&-0.027&-0.021& 0.020& \\
&138&-0.029&-0.037& 0.000& \\
&140&-0.007&-0.030& 0.000& \\
&142& 0.000& 0.000& 0 000& \\
&144& 0.000& 0.000&-0.010& \\
&146& 0.000& 0.000& 0.010& \\
&148& 0.040& 0.059& 0.030& \\
&150& 0.053& 0.067& 0.050&  \\
&152& 0.081& 0.090& 0.050&  \\
&154& 0.098& 0.113& 0.060&  \\
&156& 0.090& 0.098& 0.060&  \\
&158& 0.070& 0.082& 0.050& \\
&160& 0.055& 0.069& 0.050& \\
&162& 0.036& 0.047& 0.030&\\
&164& 0.019& 0.031& 0.030&\\
\hline\hline
\end{tabular}
\end{center}
\end{table}
\newpage
\noindent\begin{table}
\begin{center}
\bigskip
\caption{\sf The hexadecapole deformations $\beta_4$ for Gd  isotopes. See
Table 13 for details.}
\bigskip
\begin{tabular}{lll c c c c  l}
\hline\hline
& A  &    &  $\beta_4$&  &\\
\hline
&  & NL-SH  & FRDM& ETF-SI& \\
\hline
&138&-0.028&-0.036& 0.000&\\
&140&-0.012&-0.043&-0.010&\\
&142&-0.005&-0.029&-0.030&\\
&144& 0.011& 0.000& 0.000&\\
&146& 0.000& 0.000& 0.000& \\
&148& 0.015& 0.000& 0.010&\\
&150& 0.051& 0.050& 0.030&\\
&152& 0.058& 0.050& 0.050&\\
&154& 0.076& 0.073& 0.050& \\
&156& 0.083& 0.088& 0.050&\\
&158& 0.074& 0.079& 0.050&\\
&160& 0.060& 0.065& 0.050&\\
&162& 0.043& 0.043& 0.030&\\
&164& 0.022& 0.029& 0.020&\\
&166& 0.004& 0.005& 0.020&\\
\hline\hline
\end{tabular}
\end{center}
\end{table}
\newpage

\noindent\begin{table}
\begin{center}
\bigskip
\caption{\sf The hexadecapole deformations $\beta_4$ for Dy isotopes. See
Table 13 for details.}
\bigskip
\begin{tabular}{lll c c c c  l}
\hline\hline
& A  &    &  $\beta_4$&   &\\
\hline
&  &  NL-SH  & FRDM& ETF-SI&  \\
\hline
&142&-0.044&-0.049&-0.030&\\
&144&-0.017&-0.028&-0.030&\\
&146&-0.001& 0.000& 0.000&\\
&148& 0.000& 0.000& 0.000& \\
&150& 0.024& 0.000& 0.010&\\
&152& 0.046& 0.041& 0.030&\\
&154& 0.051& 0.041& 0.030&\\
&156& 0.065& 0.046& 0.040&\\
&158& 0.080& 0.060& 0.050&\\
&160& 0.070& 0.053& 0.040&\\
&162& 0.056& 0.040& 0.030&\\
&164& 0.035& 0.025& 0.020&\\
&166& 0.016& 0.010& 0.020&\\
&168&-0.001&-0.012&-0.010&\\
\hline\hline
\end{tabular}
\end{center}
\end{table}
\newpage
\noindent\begin{table}
\begin{center}
\bigskip
\caption{\sf The hexadecapole deformations $\beta_4$ for Er  isotopes. See
Table 13 for details.}
\bigskip
\begin{tabular}{lll c c c c   l}
\hline\hline
& A  &    &  $\beta_4$&  &\\
\hline
&    &NL-SH  & FRDM& ETF-SI&  \\
\hline
&142&-0.064&-0.073&-0.040&\\
&144&-0.056&-0.066&-0.030&\\
&146&-0.022&-0.035&-0.030&\\
&148&-0.025&-0.037&-0.040&\\
&150& 0.000& 0.000& 0.000&\\
&152& 0.015& 0.000& 0.010&\\
&154& 0.030& 0.040& 0.020&\\
&156& 0.033& 0.030& 0.010&\\
&158& 0.053& 0.034& 0.020&\\
&160& 0.068& 0.040& 0.010&\\
&162& 0.062& 0.037& 0.020&\\
&164& 0.050& 0.020& 0.020&\\
&166& 0.033& 0.006& 0.000&\\
&168& 0.006&-0.007&-0.001&\\
&170&-0.019&-0.023&-0.010&\\
\hline\hline
\end{tabular}
\end{center}
\end{table}
\newpage
\noindent\begin{table}
\begin{center}
\bigskip
\caption{\sf The hexadecapole deformations $\beta_4$ for Yb  isotopes. See
Table 13 for details.}
\bigskip
\begin{tabular}{lll c c c c  l}
\hline\hline
& A  &    &  $\beta_4$&    &\\
\hline
&    &NL-SH  & FRDM& ETF-SI&  \\
\hline
&154& 0.008& 0.000& 0.010&\\
&156& 0.016& 0.030& 0.010&\\
&158& 0.014& 0.034& 0.000&\\
&160& 0.015& 0.016& 0.010&\\
&162& 0.041& 0.019& 0.000&\\
&164& 0.049& 0.010& 0.010&\\
&166& 0.045& 0.003& 0.000&\\
&168& 0.030&-0.010& 0.000&\\
&170& 0.005&-0.025&-0.010&\\
&172&-0.017&-0.040&-0.030&\\
&174&-0.036&-0.059&-0.030&\\
&176&-0.059&-0.071&-0.050&\\
&178&-0.074&-0.087&-0.050&\\
&180&-0.090&-0.098&-0.080&\\
&182&-0.102&-0.117&-0.070&\\
&184&-0.084&-0.128&-0.080&\\
\hline\hline
\end{tabular}
\end{center}
\end{table}
\newpage
\noindent\begin{table}
\begin{center}
\bigskip
\caption{\sf The charge hexadecapole moment $h_{c}$ in b$^2$ for Nd isotopes
obtained in the RMF theory. The experimental data available (for details see
text) is also given for comparison.}
\bigskip
\begin{tabular}{lll c c c l}
\hline\hline
& A  & $h_{c}$$^{RMF}$&  $h_{c}$$^{expt}$  &\\
\hline
&130&0.152& - & \\
&132&0.070&-  & \\
&134&0.021&- &\\
&136&-0.013&-& \\
&138&0.003&- & \\
&140&0.000& -  & \\
&142&0.000& -  &\\
&144&0.000& -   & \\
&146&0.023& -   &\\
&148&0.099& 0.36$^{+0.10}_{-0.12}$  &\\
&150&0.376& 0.30$^{+0.06}_{-0.07}$, 0.25(12)  &\\
&152&0.510&  -  &\\
&154&0.508&  -  &\\
&156&0.465&  - &\\
&158&0.439&  - &\\
&160&0.365&  - &\\
&162&0.310&  - &\\
\hline\hline
\end{tabular}
\end{center}
\end{table}
\newpage
\noindent\begin{table}
\begin{center}
\bigskip
\caption{\sf The charge hexadecapole moments $h_{c}$  for Sm isotopes.
See Table 19 for details.}
\bigskip
\begin{tabular}{lll c  c c c cl}
\hline\hline
& A  & $h_{c}$$^{RMF}$&  $h_{c}$$^{expt}$  &\\
\hline
&134&0.114& - & \\
&136&0.064& - & \\
&138&0.021& - & \\
&140&0.027& - & \\
&142&0.000& - & \\
&144&0.000& - & \\
&146&0.000& - &  \\
&148&0.124& - & \\
&150&0.193& - &  \\
&152&0.359& 0.46(2), 0.40(9), 0.37(8)  &  \\
&154&0.468& 0.48(8), 0.63(5), 0.50$^{+0.09}_{-0.08}$ &  \\
&156&0.469&  -& \\
&158&0.432&  -& \\
&160&0.405&  -& \\
&162&0.366&  -&\\
&164&0.319&  -&\\
\hline\hline
\end{tabular}
\end{center}
\end{table}
\newpage
\noindent\begin{table}
\begin{center}
\bigskip
\caption{\sf The charge hexadecapole moments $h_{c}$   for Gd  isotopes. See
Table 19 for details.}
\bigskip
\begin{tabular}{lll c c c  l}
\hline\hline
& A  & $h_{c}$$^{RMF}$&  $h_{c}$$^{expt}$  &\\
\hline
&138&0.319&  -  &\\
&140&0.142&  -  &\\
&142&0.050&  -&\\
&144&0.055&  -  &\\
&146&0.000&   -& \\
&148&0.044&  -&\\
&150&0.185&  -& \\
&152&0.239& - &\\
&154&0.366& 0.38(16), 0.53(7), 0.64$^{+0.06}_{-0.49}$ & \\
&156&0.445& 0.42(8), 0.50(4), 0.41$^{+0.12}_{-0.18}$  & \\
&158&0.440& 0.39(9), 0.35(13), 0.34$^{+0.20}_{-0.22}$ &  \\
&160&0.419& 0.36(10), 0.35$^{+0.09}_{-0.07}$ & \\
&162&0.385&  - &\\
&164&0.339&  - &\\
&166&0.292&  - &\\
\hline\hline
\end{tabular}
\end{center}
\end{table}
\newpage
\noindent\begin{table}
\begin{center}
\bigskip
\caption{\sf The charge hexadecapole moments $h_{c}$  for Dy isotopes. See
Table 19 for details.}
\bigskip
\begin{tabular}{lll c c c   cl}
\hline\hline
& A  & $h_{c}$$^{RMF}$&  $h_{c}$$^{expt}$  &\\
\hline
&142&-0.001& - &\\
&144&0.028&  -&\\
&146&0.012&  -& \\
&148&0.000&  -& \\
&150&0.082&  -   & \\
&152&0.185&   - &  \\
&154&0.228&   - &  \\
&156&0.313&  0.21$^{+0.16}_{-0.20}$  &  \\
&158&0.419&  0.16 $^{+0.10}_{-0.15}$  &  \\
&160&0.407&   - &  \\
&162&0.392&  0.27(10) &  \\
&164&0.338&  0.28(10), 0.25(16), 0.23$^{+0.10}_{-0.12}$  & \\
&166&0.313&  -   & \\
&168&0.252&  -   & \\
\hline\hline
\end{tabular}
\end{center}
\end{table}
\newpage
\noindent\begin{table}
\begin{center}
\bigskip
\caption{\sf The charge hexadecapole moments $h_{c}$ for Er  isotopes. See
Table 19 for details.}
\bigskip
\begin{tabular}{lll c c c  l}
\hline\hline
& A  & $h_{c}$$^{RMF}$&  $h_{c}$$^{expt}$  &\\
\hline
&142&-0.050& -&\\
&144&-0.053& -&\\
&146&0.018&  -&\\
&148&-0.011& -&\\
&150&0.000&  -&\\
&152&0.061&  -&\\
&154&0.134&  -&\\
&156&0.164&  -& \\
&158&0.271&  - &\\
&160&0.366&  - &\\
&162&0.377& 0.16$^{+0.14}_{-0.26}$  &\\
&164&0.361& 0.12$^{+0.12}_{-0.13}$ &\\
&166&0.323& 0.30(2), 0.32(16), 0.22$^{+0.11}_{-0.16}$ &\\
&168&0.248& 0.18(2), 0.20$^{+0.12}_{-0.18}$ &\\
&170&0.165& 0.31(2), 0.24$^{+0.14}_{-0.18}$ &\\
\hline\hline
\end{tabular}
\end{center}
\end{table}
\newpage
\noindent\begin{table}
\begin{center}
\bigskip
\caption{\sf The charge hexadecapole moments $h_{c}$ for Yb  isotopes. See
Table 19  for details.}
\bigskip
\begin{tabular}{lll c c c  l}
\hline\hline
& A  & $h_{c}$$^{RMF}$&  $h_{c}$$^{expt}$  &\\
\hline
&154&0.042&  -&\\
&156&0.086&  -& \\
&158&0.096&  - &\\
&160&0.122&   -&\\
&162&0.244&   -&\\
&164&0.324&  -&\\
&166&0.349&  - &\\
&168&0.307& 0.19$^{+0.14}_{-0.19}$, +0.10$^{+0.10}_{-0.09}$ &\\
&170&0.214&  - &\\
&172&0.130&  0.22$^{+0.12}_{-0.18}$ & \\
&174&0.058&  0.23(17), 0.22$^{+0.14}_{-0.18}$ & \\
&176&-0.029&  - & \\
&178&-0.086&  - &\\
&180&-0.149&  - &\\
&182&-0.196&  - &\\
&184&-0.162&  - &\\
\hline\hline
\end{tabular}
\end{center}
\end{table}

\end{document}